\documentclass[preprint,12pt]{elsarticle}

\usepackage{amssymb}

\usepackage{graphicx}
\usepackage{amsmath,epsfig}
\usepackage{natbib}
\def\be{\begin{equation}}
\def\ee{\end{equation}}
\def\bea{\begin{aligned}}
\def\eea{\end{aligned}}
\def\ba{\begin{eqnarray}}
\def\ea{\end{eqnarray}}

\def\yzero{\smash{\hbox{$y\kern-4pt\raise1pt\hbox{${}^\circ$}$}}}

\def\-{\hphantom{-}}

\def\s2{\frac{1}{\sqrt2}}

\def\IF{\relax{\rm I\kern-.18em F}}
\def\II{\relax{\rm I\kern-.18em I}}
\def\IP{\relax{\rm I\kern-.18em P}}
\def\IC{\relax\hbox{\kern.25em$\inbar\kern-.3em{\rm C}$}}
\def\IR{\relax{\rm I\kern-.18em R}}

\def\Dsl{\,\raise.15ex\hbox{/}\mkern-13.5mu D} 
\def\IZ{Z\kern-.4em  Z}

\journal{Physics Letter B}

\begin{document}

\begin{frontmatter}



\title{M2-branes on a constant flux background}
\author[label a]{M.P. Garcia del Moral}
\author[label a]{C. Las Heras}
\author[label a]{P. Leon}
\author[label b]{J.M. Pena}
\author[label a]{A. Restuccia}
\address[label a]{Departamento de F\'isica, Universidad de Antofagasta, Aptdo 02800, Chile.}
\address[label b]{Departamento de F\'isica,  Facultad de Ciencias,
Universidad Central de Venezuela, A.P. 47270, Caracas 1041-A, Venezuela\footnote{This work started to be written in a Visiting Research (August-September 2018) at Departamento de F\'\i sica, Universidad de Los Andes, Bogot\'a, Colombia.}}




\begin{abstract}
We describe a compactified Supermembrane, or M2-brane,  with 2-form fluxes generated by  constant three-forms  that are  turned on a 2-torus of the target space $M_9\times T^2$. We compare this theory with the one describing a $11D$ M2-brane formulated on $M_9\times T^2$ target space subject to an irreducible wrapping condition.  We show that the flux generated by the bosonic 3-form under consideration is in a one to one correspondence to the irreducible wrapping condition. After a canonical transformation both Hamiltonians are exactly the same up to a constant shift in one particular case. Consequently both of them, share the same spectral properties. We conclude that the Hamiltonian of the M2-brane with 2-form target space fluxes on a torus has a purely discrete spectrum with eigenvalues of finite multiplicity and it can be considered to describe a new sector of the microscopic degrees of freedom of M-theory. We also show that the total membrane momentum in the direction associated to the flux condition adquires a quantized contribution in correspondence to the flux units that have been turned on.

\end{abstract}


\end{frontmatter}



\section{Introduction} 
Supermembranes, also called M2-branes, are $2+1$ supersymmetric extended objects that evolve in an eleven dimensional target space \cite{Bergshoeff}. They are described by  a nonlinear interacting field theory invariant under global supersymmetry, local diffeomorphisms and local kappa symmetry \cite{dwhn}. Supermembranes are part of the building blocks of M-theory and they are sources of $11D$ supergravity \cite{Bergshoeff,Bergshoeff2}, a relation that was emphasized in \cite{Ovrut}.  Moreover, in 
\cite{Duff2} it is shown that the supermembrane emerges as an exact solution of supergravity field equations.
A remarkable property of supermembrane theory is that all the five string theories at least at kinematical level and by double dimensional reduction can be obtained from it \cite{Aldabe:1998ie,Aldabe2,Schwarz3,Duff}. Recently it has been shown that M2-brane toroidally compactified is U-dual invariant \cite{MPGM8,MPGM10}. 

The Supermembrane theory  was  originally expected to describe the microscopic degrees of freedom of M- theory, however when formulated on 11D  Minkowski background \cite{dwhn}, it was rigorously proved in the context of matrix model regularization  that it has continuous spectrum from $[0,\infty)$ \cite{dwln}, and the toroidal compactification by itself does not change this behaviour \cite{dwpp}. This property led to the community to re-interpret this theory as a second quantized theory  \cite{bfss}.  There are just two cases described in the literature -up to our knowledge- in which the formulation of supermembrane theory exhibits discreteness of the spectrum \cite{bgmr,bgmr3}. A first case the so called \textit{supermembrane  with central charges} \cite{MOR} that is irreducibly wrapped \cite{MRT} around a 2-torus, -condition that has been extended to other compactifications, see for example \cite{bellorin-restuccia,MPGM11,belhaj}, and a second one, corresponding the supermembrane on a pp-wave background \cite{Sugiyama, Dasgupta} whose matrix model regularization corresponds to the BMN matrix model \cite{BMN} and whose properties of discreteness were proven on \cite{bgmr3}.

Since M-theory is a candidate for unification theory, at least a sector of the theory will be described in terms of the Supermembrane theory degrees of freedom. Consequently it becomes increasingly clear the need to obtain the M2-brane theory formulated on more general backgrounds. Previous formulation of the M2-brane in the Light Cone Gauge (L.C.G) on arbitrary curved backgrounds in the formalism of the superspace to second order in grassmann variables was done in \cite{dwpp3}.
In this paper we analise the quantum properties of a simple but nontrivial case of a  M2-brane  on  a $M_9\times T^2$ background with particular target space 2-form fluxes generated by the presence of non vanishing components of a constant form $C_{\pm}$. These backgrounds are consistent with supergravity. Indeed, backgrounds with a constant  $C_3$ were considered by the authors \cite{Duff2}, corresponding to a M2-brane acting as a source. The analysis that we do includes backgrounds corresponding to the asymptotic limit of those considered in \cite{stelle-pbranes}. 
The paper is organized as follows: In section 2 we obtain the Hamiltonian and constraints of the M2-brane on $M_9\times T^2$ with target space  fluxes. In section 3 we compare this theory with the a M2-brane theory irreducibly wrapped around the 2-torus.  Based on this result we characterize its spectrum. In section 4 we establish the relation between the two formulations and based on this result we characterize its spectral properties. In section 5 we present our conclusions.

\section{M2-brane subject to 2-form fluxes induced by  constant $C_3$}
In this section we will analyse  the supermembrane theory in the L.C.G. formulated on $M_9\times T^2$ background  with 2-form fluxes induced by the presence of constant bosonic 3-form gauge fields $C_{\mu \nu \lambda }$. The supersymmetric action of the M2-brane on a generic 11D noncompact background was found by 
\cite{Bergshoeff}. 
\begin{equation}
\small
\label{supermembrane}
S= T \int d^3\xi\biggl\lbrack -\frac{1}{2}{\sqrt{-g}}{g}^{uv}
\Pi_u{}^{{{\hat{m}}}}\Pi_v{}^{{\hat{n}}} \eta_{{{\hat{m}}}{\hat{n}}}+\frac{1}{2}{\sqrt{-g}}
-\frac{1}{6}\varepsilon ^{uvw}\Pi_{u}^{{\hat{A}}}
\Pi_{v}^{{\hat{B}}} \Pi_{w}^{{\hat{C}}}C_{{\hat{C}}{\hat{B}}{\hat{A}}}
\biggr\rbrack \,,
\end{equation}
where, $g_{uv}=\Pi_u{}^{\hat{m}}\Pi_v{}^{{\hat{n}}} \eta_{{{\hat{m}}}{\hat{n}}}$ with $u,v,w,=0,1,2$ the worldvolume indices. (Here $\hat{A}, \hat{B}, \hat{C},$ are tangent superspace indices; ${\hat{m}},{\hat{n}},{\hat{l}},$ bosonic tangent space indices and $\hat{a}, \hat{b}, \hat{c},$ fermionic tangent space indices).  Pullback of the supervielbein $E_M^{\hat{A}}$ to the worldvolume  is given by $\Pi_u^{\hat{A}}=\frac{\partial Z^M}{\partial \xi^u}E_M^{\hat{A}}$ where $M$ is superspace index.
In our study we will focus on flat metric $G_{\mu \nu} =\eta_{\mu \nu}$ backgrounds but in the presence of some constant components of the three-form:
\begin{equation}
\begin{aligned}
\small
\label{background}
&\Pi_v^{\hat{m}} = \partial_v X^\mu \delta_{\mu}^{\hat{m}}  +\bar{\theta}\Gamma^{\hat{m}} \partial_u\theta , \qquad \Pi_v^{\hat{a}}=\partial_v\theta^{\hat{a}} \,, \\
&C_{\mu \alpha \beta}=(\bar{\theta}\Gamma_{\mu \nu})_{(\alpha}(\bar{\theta}\Gamma^\nu)_{\beta )}\quad,  \quad \, \, C_{\alpha \beta \gamma}=(\bar{\theta}\Gamma_{\mu \nu})_{(\alpha}(\bar{\theta}\Gamma^\mu)_{\beta}(\bar{\theta}\Gamma^\nu)_{\gamma)} \quad, \\
&C_{\mu \nu \rho}= const \quad ,\qquad \qquad \quad \quad C_{\mu \nu \alpha}=(\bar{\theta}\Gamma_{\mu\nu})_{\alpha} \quad,
\end{aligned}
\end{equation}
\noindent  The embedding coordinates in the superspace formalism are $(X^{\mu}(\xi),\theta^{\alpha}(\xi))$ with $\xi^u $ the worldvolume coordinates and where $\mu, \nu, \lambda $ and  $\alpha, \beta,$ are bosonic and fermionic target space indices, respectively. They are scalars under reparametrizations on $\xi$.
In this background the action of the supermembrane takes the following form:
\begin{equation}
\small
\begin{aligned}
\label{2.6}
S= & - T \int d^3 \xi \{ \sqrt{-g}+\varepsilon^{uvw}\bar{\theta}\Gamma_{\mu \nu}\partial_w \theta \left[ \frac{1}{2}\partial_u X^\mu (\partial_v X^\nu\right. +  \bar{\theta}\Gamma^\nu \partial_v\theta) + \\
& + \frac{1}{6}\bar{\theta}\Gamma^\mu \partial_u \theta \bar{\theta}\Gamma^\nu \partial_v \theta ] +\frac{1}{6}\varepsilon^{uvw}\partial_u X^\mu \partial_v X^{\nu} \partial_w X^\rho C_{\rho \nu \mu} \} \, .
\end{aligned}
\end{equation}

This background is consistent with supergravity in 11D dimensions and in a particular case it corresponds to the asymptotic limit of a supergravity solution generated by an M2-brane acting as a source \cite{Duff2, stelle-pbranes}. The line element in this case is given by 
\begin{equation}
\small
\label{metric-stelle}
ds^2=(1+\frac{k}{r^6})^{-\frac{2}{3}}{dx^{\bar{\mu}}}{dx^{\bar{{\nu}}}}{\eta_{{\bar{\mu}} {\bar{\nu}}}} + (1+\frac{k}{r^6})^{-\frac{1}{3}}{dy^{\bar{m}}}{dy^{\bar{{n}}}}{\delta_{{\bar{m}} {\bar{n}}}} \, ,
\end{equation}
where ${\bar{\mu}}= 0, 1, 2$, ${\bar{{m}}} = 3, ... , 10$ and $r = \sqrt{y^{\bar{{m}}}y^{\bar{{m}}}}$ is the radial isotropic coordinate in the transverse space. On the other hand, the ansatz for the 3-form produces:
\begin{equation}
\small
\label{3form-stelle}
C_{{\bar{\mu}}{\bar{\nu}}{\bar{\sigma}}} = \epsilon_{{\bar{\mu}}{\bar{\nu}}{\bar{\sigma}}} (1+\frac{k}{r^6})^{-1} \, ,
\end{equation}
with the other components set to zero. When $r\rightarrow \infty $, the metric (\ref{metric-stelle}) goes to Minkowski metric and (\ref{3form-stelle}) is constant.
We may now formulate the supermembrane action on a $M_9 \times T^2$ target space with constant gauge field $C_{{{\mu}}{{\nu}}{{\sigma}}}$ closely following the definitions of \cite{dwpp}. We start the L.C.G fixing by taking,
$X^+(\xi)= X^{+}(0)+ \tau $, where $\tau$ is the time coordinate on the worldvolume, so that ${\partial}_u X^{+}= {\delta}_{u\tau}$, and ${\Gamma}^{+} \theta = 0$. 
Now the supersymmetric action formulated in this partial gauge fixing is
\begin{equation}
\small
S = T\int d^3 \xi \{ -  \sqrt{\bar{g}\Delta}-\varepsilon^{rs}\partial_rX^a \bar{\theta} \Gamma^- \Gamma_a \partial_s \theta +  C_+ +\partial_{\tau} X^- C_- +\partial_{\tau} X^a C_a+C_{+-} \}
\end{equation}
with\footnote{In order to be self-contained we include the definitions 
of  \cite{dwhn}:
$\Delta = -g_{00} + u_r {\bar{g}}^{rs} u_s$ being ${\bar{g}}^{rs} g_{st} = {{\delta}^r}_t$ and $g \equiv det g = - \Delta \bar{g}$  (with $\varepsilon^{0rs}=\varepsilon^{rs}$).
$
{\bar{g}}_{rs} \equiv g_{rs} = \partial_r X^a \partial_s X^b {\delta}_{ab} ;\quad
$
$u_r \equiv g_{0r} = \partial_r X^- + \partial_0 X^a \partial_r X^b {\delta}_{ab} + \bar{\theta}{\Gamma}^{-} \partial_r \theta ;$
${\bar{g}}_{\small 00} = 2\partial_0 X^- + \partial_0 X^a \partial_0 X^b {\delta}_{ab} + 2\bar{\theta}{\Gamma}^{-} \partial_0 \theta.
$}
\begin{equation}
\small
\begin{aligned}
& C_a  =  -\varepsilon^{rs}\partial_rX^- \partial_sX^b C_{-ab} +\frac{1}{2}\varepsilon^{rs}\partial_rX^b \partial_sX^c C_{abc} \, , \\
& C_{\pm}  =  \frac{1}{2}\varepsilon^{rs}\partial_rX^a \partial_sX^b C_{\pm ab} \,, \qquad C_{+-}  =  \varepsilon^{rs}\partial_rX^- \partial_sX^a C_{+-a} \,,
\end{aligned}
\end{equation}
where $\sigma^r$, $r=1,2$ are the spacelike coordinates of the base manifold $\Sigma \times R$, being $\Sigma$ a torus. It is possible to fix the variation of some components of the 3-form by virtue of its gauge invariance. In particular it is possible to fix $C_{+-a}=0$ and $C_{- ab}=constant.$ 
The Hamiltonian takes the form
\begin{equation}
\small
H=T\int{d^2 \sigma}\{\frac{1}{(P_ -- C_-)}\left[\frac{1}{2}(P_a-C_a)^2  +\frac{1}{4}\left(\varepsilon^{rs}\partial_r X^a\partial_s X^b\right)^2\right] + \varepsilon^{rs}\bar{\theta}\Gamma^- \Gamma_a \partial_s \theta \partial_r X^a -C_{+}\}\,
\end{equation}
subject to the primary constraints
%
\begin{equation}
\small
\label{ecuacion10constraints}
P_a \partial_r X^a + P_-\partial_r X^- + \bar{S}\partial_r \theta \approx  0 \quad ,  \quad
S+(P_- -C_-)\Gamma^{-}\theta \approx  0\,.
\end{equation}
with 
\begin{equation}
\small
P_a  =  \sqrt{\frac{\bar{g}}{\Delta}}(\partial_{\tau}X_a -u_r \bar{g}^{rs}\partial_s X_a)+C_a,\quad
P_-  =  \sqrt{\frac{\bar{g}}{\Delta}}+C_-,\quad
S  = -\sqrt{\frac{\bar{g}}{\Delta}} \Gamma^- \theta. \label{cm6}
\end{equation}
The contribution of $C_{+}$ to the Hamiltonian is a boundary term that does not necessarily vanish since there exists a compact sector of target space. $X^-$ appears explicitly through $C_a$ in the Hamiltonian. On the other side $X^-$ can be solved from the constraint in terms of the physical degrees of freedom. However non-locality is introduced in the procedure.
In order to achieve a local canonical polynomial reduction of the Hamiltonian one may perform a the following transformation
\begin{equation}
\small
\label{cambio}
P_a \rightarrow \hat{P}_a \equiv  P_a -C_a ,  \qquad P_- \rightarrow \hat{P}_- \equiv P_- -C_-\,,
\qquad S \rightarrow \hat{S} \equiv S \,,
\end{equation}
keeping invariant the rest of the canonical variables.
These preserve all the Poisson brackets, as it is a canonical transformation on the phase space. In fact, the kinetic terms remain invariant under (\ref{cambio}),
\begin{equation}
\small
\int_{\Sigma} (P_a \dot{X}^a +P_{-}\dot{X}^{-}+ \bar{S} \dot{\theta} ) = \int_{\Sigma} ({\hat{P}}_a \dot{\hat{X}}^a +{\hat{P}}_{-}\dot{\hat{X}}^{-}+ \hat{\bar{S}} \dot{\hat{\theta}})\,.
\end{equation}
and the new constrains are
\begin{eqnarray}
\small
P_a \partial_r X^a + P_-\partial_r X^- + \bar{S}\partial_r \theta  & = & \hat{P}_a \partial_r \hat{X}^a + \hat{P}_-\partial_r \hat{X}^- + \bar{ \hat{S}}\partial_r \hat{\theta} \approx 0 \\
\chi \approx S+(P_- -C_-)\Gamma^{-}\theta & = &  \hat{S}+\hat{P}_-\Gamma^{-}\hat{\theta} \approx 0
\end{eqnarray}
Now we may use the residual gauge symmetry 
generated by the constraints to impose the gauge fixing condition
$
{\hat{P}}_{-}= \hat{P}_{-}^0 \sqrt{w} \,,
$
\noindent where $\sqrt{w}$ is a time independent scalar density and $\hat{P}_{-}^0$ a zero mode defined as in \cite{dwhn}.
We may then eliminate $(\hat{X}^-,{\hat{P}}_{-}) $ as canonical variables and obtain a formulation solely in terms of $(\hat{X}^a, \hat{P}_a)$. The remaining constraint after the partial gauge fixing corresponds, as usual, to the area preserving ones:
\begin{equation}
\small
\label{ecuacionVII}
d({\hat{P}_{a}}d\hat{X}^a+\overline{\hat{\theta}}\Gamma^{-}d\hat{\theta}) = 0 \,, \qquad
\int_{\Sigma} d({\hat{P}}_{a} d \hat{X}^a+\overline{\hat{\theta}}\Gamma^{-}d\hat{\theta})  = 0 \, ,
\end{equation}
where the first constraint is the local integrability condition which must be satisfied in order to have a solution for $\hat{X}^{-}$. The second integral constraint, is the condition that the periods $d\hat{X}^{-}$ are trivial and hence $d\hat{X}^{-}$ is an exact one form.
We then have the following hamiltonian density for a membrane on a general background field, after the elimination of the conjugate pairs $(\hat{X}^{-}, \hat{P}_{-})$ and $(\hat{X}^{+}, \hat{P}_{+})$,
\begin{equation}
\small
\label{hamiltoniantecho}
{\hat{\mathcal{H}}}_{total}=\hat{\mathcal{H}}-C_{+}= \frac{1}{\hat{P}_-}\left[\frac{1}{2}\hat{P}_a\hat{P}^a +\frac{1}{4}\left(\varepsilon^{rs}\partial_r \hat{X}^a\partial_s \hat{X}^b\right)^2 \right]+\varepsilon^{rs}\bar{\hat{\theta}}\Gamma^- \Gamma_a \partial_s \hat{\theta} \partial_r \hat{X}^a -C_{+}
\end{equation}
Since we are considering a toroidal compactification of the target space $M_{9}\times T^2$, the bosonic components decompose in the compact and non-compact sector. The  even embedding maps $\hat{X}^{a}$ associated to the compact sector which always appear as closed one-forms in the action. They decompose into an exact one-form, plus a harmonic one-form. The latter may have nontrivial periods on the basis of homology of the compact base Riemann surface $\Sigma$. The odd embedding maps $\hat{\theta}^{\alpha}$ we assume to be single valued on the base manifold.  
The gauge fixing procedure is consistent under this compactification since the local gauge transformations only involve the exact part of $\hat{X}^{a}$. The Hamiltonian of the compactified theory is the following one:
\begin{equation}
\small\small
\label{hamiltonian-fluxes}
\begin{aligned} 
\int_{\Sigma}d^2\sigma{\hat{\mathcal{H}}}_{total} = \int_{\Sigma} d^2\sigma \{ \frac{\sqrt{w}}{\hat{P}_{-}^0} \, \left[ \frac{1}{2} \left(\frac{P_m}{\sqrt{w}} \right)^2 + \frac{1}{2}\left(\frac{P_i}{\sqrt{w}} \right)^2 + \frac{1}{4} \left\{ X^i, X^j \right\} ^2 +  \frac{1}{2} \left\{ X^i, X^m \right\}^2  \right. \\
\left. + \frac{1}{4} \left\{ X^m, X^n \right\} ^2 \right] +\sqrt{w} \left[ \bar{\theta}\Gamma^-\Gamma_m\left\lbrace X^m,\theta\right\rbrace +\bar{\theta}\Gamma^-\Gamma_i\left\lbrace X^i,\theta\right\rbrace\right] \} -C_+\,,
\end{aligned}
\end{equation}
where the index $m$ denotes the maps from the base $\Sigma$ to $M_9$ and $i,j=1,2$ the map  from $\Sigma$ to $T^2$. Generically, $dX^i= M^i_j d\hat{X}^j + dA^i$, $M^i_j$ are integers in order to have a map to cycles. $d\hat{X}^j$ , is a normalized basis of harmonic one-forms and $dA^i$ are exact one-form components. There are no further requirements on $M^i_j$.
We notice the differences between the above Hamiltonian and the one on a Minkowski $M_{11}$ target space where no harmonic contribution is present. We will give the complete expression of the Hamiltonian in terms of $d\hat{X}^i, A^i, dX^m$ in the next section. This Hamiltonian is subject to the local and global constraints associated to the Area Preserving Diffeomorphisms
\begin{equation}
\small
\label{constraintdos}
d(P_i dX^i + P_m dX^m+\overline{\theta}\Gamma^{-}d\theta)=0 \,, \qquad
 \oint_{{\mathcal{C}}_s} d(P_i dX^i + P_m dX^m+\overline{\theta}\Gamma^{-}d\theta)=0 \,.
\end{equation}
In the compactified case, in contrast to the noncompact one, the last term in (\ref{hamiltonian-fluxes}) for constant bosonic 3-form is a total derivative of a multivalued function (due to the harmonic contribution). Therefore its integral is not necessarily zero.

Classically the dynamics of this Hamiltonian contains string-like spikes which render the quantum spectrum of the theory continuous.
The main point to study in this paper is the behaviour of the theory when we add a flux quantization condition on the 3-form.  Given the target space $M_9 \times T^2$ a flux condition on it corresponds to a closed two form $F_2$ whose integral on the compact sector is an integer number. This flux condition is equivalent to the existence of an $U(1)$ principle bundle over $T^2$ and of a 1-form connection on it whose curvature is $F_2$:
\begin{equation}
\small
\label{flux condition}
\int_{T^2}{F_2}
= k \in \mathbb{Z}/\{0\} \,.
\end{equation}
In this paper we consider the closed two-forms generated by $C_{+}$ or $C_{-}$. We will define them in the following sections.
We are interested in the quantum properties of the supermembrane on a target space with nontrivial $C_{\pm}$, under a flux condition generated by them. In order to perform this analysis we are going to compare the Hamiltonian (\ref{hamiltonian-fluxes}) subject to a flux condition with the Hamiltonian of the M2-brane irreducibly wrapped on a target space $M_9 \times T^2$.
%
%
%

\section{The M2-brane with irreducible wrapping}
The M2-brane with irreducible wrapping is defined as follows. The embedding maps satisfy winding conditions over the nontrivial 1-cycles of the 2-torus with  $\oint_{C_j} dX^i=M^i_j$, $M^i_j$ are winding numbers integers and $C_s$ the homological basis of the 2-torus. The embedding maps associated to the wrapping on $T_2$ satisfy the following topological condition \cite{MRT}
\begin{equation}
\small
\label{topologicalcondition}
\int_{\Sigma} dX^i \wedge dX^j =\epsilon^{ij} n A,\quad n\in \mathbb{Z}/\{0\} \,,
\end{equation}
with $A$ denoting the area of the 2-torus (which we can be normalized to 1) and $n$ an integer that is chosen to be different from zero.  This condition is a quantization condition and ensures that the harmonic modes appear in a nontrivial way in the expression of $X^i.$ This condition is related with the existence of a central charge in the SUSY algebra  and for this reason this sector of the Supermembrane has been denoted as \textit{Supermembrane with central charges}. Indeed, it implies that the supermembrane is a calibrated submanifold \cite{bellorin-restuccia}. Other studies analyzing  the  M2-brane on holomorphic curves was considered in \cite{Husain:2002tk} and \cite{Gauntlett:2001qs}. From a geometrical point of view, irreducibility condition ensures the existence of a nontrivial $U(1)$ principal bundle over the worldvolume of the supermembrane, characterized by the integer $n$ associated to its first Chern class. A particular $n$ fixes and restricts the allowed  class of principal fiber bundle where it can be formulated.
The canonical connections are $U(1)$ monopoles expressed in terms of the embedding maps (which are minimal immersions) of the supermembrane in the compactified space. Indeed, this  corresponds to have a nontrivial 2-form flux over the supermembrane world-volume
\begin{equation}
\small
\int_{\Sigma}F_2= n\in\mathbb{Z}/\{0\} \,.
\end{equation}
\label{flux}
See \cite{MRT} for further details. The Hamiltonian of a supermembrane wrapped on a 2-torus subject to (\ref{topologicalcondition}) found in \cite{MOR} is the following one
\begin{equation}
\small
\begin{aligned}
\label{hamiltonianirred}
H^{\tiny{Irred}}&=\int_\Sigma d^2\sigma\sqrt{w}\Big[\frac{1}{2}\Big(\frac{P_m}{\sqrt{w}}\Big)^2+\frac{1}{2}\Big(\frac{P_i}{\sqrt{w}}\Big)^2 + \frac{1}{4}\left\{X^m,X^m\right\}^2 + \frac{1}{2}(\mathcal{D}_iX^m)^2+\frac{1}{4}(\mathcal{F}_{ij})^2 \Big] \\
&+ \int_\Sigma d^2\sigma\sqrt{w}\Big[\Lambda\Big(\mathcal{D}_i\big(\frac{P_i}{\sqrt{w}}\big)+\left\{X^m,\frac{P_m}{\sqrt{w}}\right\} \Big)\Big] +(n^2Area_{T^2}^2)\\
&+ \int_\Sigma d^2\sigma\sqrt{w}\Big[-\bar{\theta}\Gamma_-\Gamma_i\mathcal{D}_i\theta-\bar{\theta}\Gamma_-\Gamma_m\left\{X^m,\theta\right\}+\Lambda\left\{\bar{\theta}\Gamma_-,\theta\right\}\Big]\,,
\end{aligned}
\end{equation}
where there is a symplectic covariant derivative and symplectic curvature defined
\begin{equation}
\small
\mathcal{D}_iX^m = D_iX^m+\left\{ A_i,X^m\right\}, \qquad  \mathcal{F}_{ij}= D_iA_j-D_jA_i+\left\{ A_i,A_j\right\}, \,
\end{equation}
with $D_i$ a covariant derivative defined in terms of the moduli of the torus, the winding numbers $M^i_j$ and the harmonic one-forms. see \cite{MPGM8}. The symplectic connection transforming under area preserving diffeomorphisms given by
$\delta_{\epsilon}A=\mathcal{D}\epsilon.$ 
In \cite{MPGM}, the authors showed that this hamiltonian classically does not contain string-like configurations. At a quantum level it has  the remarkable property of having a supersymmetric discrete spectrum with finite multiplicity, \cite{bgmr, bgmr3} in distinction with the supermembrane compactified on a torus without this restriction (\ref{topologicalcondition}) which has continuous spectrum from $[0, \infty)$ \cite{dwln,dwpp}. The irreducible wrapping condition is  a flux condition over the worldvolume that generalizes the Dirac monopole construction to Riemann surfaces of arbitrary genus $\ge 1$ \cite{MR}. The theory defined in this way is a restriction of the supermembrane theory. All configurations must satisfy, in addition, the global constraint. The constraint (\ref{topologicalcondition}) does not change the local symmetries of the supermembrane theory since it is topological condition. In particular, the invariance under area preserving diffeomorphisms is preserved.
%
%
\section{Relation between both formulations}
We start by considering a flux condition generated by $C_{\pm}$ on the base manifold $\Sigma$. $C_{\pm}$ is a density with the dimensions of the membrane momentum in the directions associated with coordinates $X^{\pm}$. It is defined on $\Sigma$, we then consider its associated 2-form 
\begin{equation}
\small
C_{\pm}d\sigma^1 \wedge d\sigma^2=\frac{1}{2}\frac{\partial X^a}{\partial \sigma^r} \frac{\partial X^b}{\partial \sigma^s}C_{\pm a b} \, d\sigma^r \wedge d\sigma^s = \frac{1}{2}C_{\pm a b}\, dX^a(\sigma, \tau) \wedge dX^b(\sigma, \tau) 
\end{equation}
Under the assumption that $C_{\pm a b}$ is constant, it is a closed two-form on $\Sigma$. We impose the flux condition
\begin{equation}
\small
\label{fluxcondition2}
\int_{\Sigma}F_{2}=\int_{\Sigma}C_{\pm}d\sigma^1 \wedge d\sigma^2 = k_{\pm} \in\mathbb{Z}/\{0\} \,.
\end{equation}
The maps from $\Sigma \rightarrow M_9 \times T^2$ decompose into maps from $\Sigma \rightarrow M_9$ and the ones from $\Sigma \rightarrow T^2$. The former are labeled with an index $m$ and the latter with index $i,j$. As we have stated the closed one-forms $dX^i$ can always be expressed in terms of the harmonic part $M^i_j d\hat{X}^j$ and its exact part. In the flux condition the exact one-forms cancel and we are only left with the harmonic sector. We then have for the flux condition 
\begin{equation}
\small
\label{flux condition}
\int_{\Sigma}F_{2}= \int_{\Sigma}\frac{1}{2}{C_{\pm ij}} M^i_k M^j_l d\hat{X}^k\wedge d \hat{X}^l=  k_{\pm} \in\mathbb{Z}/\{0\} \,,
\end{equation}
 The normalized basis of harmonic one-forms on $\Sigma$, $dX^i$, satisfy
$
\int_{C_i}{d \hat{X}^j}= \delta^j_i \,,
$
with 
$C_i$, being the homology basis on $\Sigma$. Using the bilinear Riemann relations we get
%
%
$\int_{\Sigma} d\hat{X}^1\wedge d \hat{X}^2= 1$
%
%
It is convenient to define the density ${\sqrt{w}}$ introduced in the gauge fixing procedure as
%
%
${\sqrt{w}}=\frac{\partial \hat{X}^1}{\partial \sigma^r} \frac{\partial \hat{X}^2}{\partial \sigma^s} \epsilon^{rs}$ 
%
%
it is a regular density on $\Sigma$, we then have
$
\int_{\Sigma}{\sqrt{w}} \, d\sigma^1 \wedge d\sigma^2= 1 \,.
$
We can now change variables from $(\sigma^1,\sigma^2)$ on $\Sigma$ to local coordinates $(\widetilde{X}^1,\widetilde{X}^2)$ on the 2-torus $T^2$, the map is defined by
$
\widetilde{X}^i = \hat{X}^i(\sigma^1,\sigma^2 ) \,.
$
In fact, the Jacobian of the change of variables is ${\sqrt{w}}$ which is nonzero on $\Sigma$. $C_{\pm}$ defines then a flux condition on $T^2$,
\begin{equation}
\small
\label{fluxcondition2}
\int_{\Sigma}F_{2}= {c_{\pm}} det M \int_{T^2} d\widetilde{X}^1\wedge d \widetilde{X}^2= \int_{T_2}\frac{1}{2}{C_{\pm ij}} M^i_k M^j_l d\widetilde{X}^k \wedge d \widetilde{X}^l \,=\int_{T^2}\widetilde{F}_2.
\end{equation}
where we have expressed $C_{\pm ij} = {c_{\pm}} {\epsilon_{ij}}$, and denoted $det M $ the determinant of the matrix $ M^i_j$.
There is then a one to one correspondence between the flux condition generated by ${C_{\pm}}$ on $\Sigma$ and on $T^2$. The main point is that $det M$ must be nonzero, that is the condition of irreducible wrapping must be satisfied. There is then a one to one correspondence between the supermembrane with irreducible wrapping and the supermembrane on a background with a flux condition on $T^2$ generated by ${C_{\pm}}$. In fact, given the latter it implies that the supermembrane has an irreducible wrapping. Conversely given a supermembrane with irreducible wrapping there always exists a three form with a flux condition compatible with the nontrivial wrapping. Moreover, if there is no flux, (\ref{fluxcondition2}) is equal to zero, then  supermembranes with reducible wrapping are admissible in the configurations space and the spectrum is consequently continuous from zero to infinity.

Furthermore the Hamiltonian of both theories differ at most in a constant, arising from the ${C_{+}}$ term in the Hamiltonian (\ref{hamiltonian-fluxes}), hence the spectrum of the supermembrane with fluxes generated by ${C_{\pm}}$ has also discrete spectrum with finite multiplicity, a remarkable property. The effect of the ${C_{\pm}}$ background produces a discrete shift in some components of the momentum of the supermembrane, and in the Hamiltonian density. Comparing with the original configuration variables $(X^a,P_a)$ and considering the total momentum of the supermembrane, we have
\begin{equation}
\small
P^0_{-}=\int_{\Sigma}P_{-} d\sigma^1 \wedge d\sigma^2=\int_{\Sigma}(\hat{P}_{-}+C_{-}) d\sigma^1 \wedge d\sigma^2 = \hat{P}^0_{-}+ k_-\,,
\end{equation}
\begin{equation}
\small
P^0_{+}=\int_{\Sigma}P_{+} d\sigma^1 \wedge d\sigma^2=\int_{\Sigma}(\hat{P}_{+}+C_{+}) d\sigma^1 \wedge d\sigma^2 = \int_{\Sigma}\hat{\mathcal{H}} d\sigma^1 \wedge d\sigma^2 + k_+\,, 
\end{equation}
\begin{equation}
\small
{P^0_{a}}=\int_{\Sigma}{P_{a}} d\sigma^1 \wedge d\sigma^2=\int_{\Sigma}(\hat{P}_{a}+C_{a}) d\sigma^1 \wedge d\sigma^2 = \int_{\Sigma}\hat{{P}_a} d\sigma^1 \wedge d\sigma^2 \,,
\end{equation}
where we have used (\ref{fluxcondition2}) and 
$
\int_{\Sigma}C_{a}d\sigma^1 \wedge d\sigma^2 = 0 \,.
$ $\hat{\mathcal{H}}$ is the Hamiltonian density of the supermembrane with fluxes (analogously with irreducible wrapping). We conclude that the interaction of the supermembrane with the background we have considered has render a quantized change of the membrane momentum on the directions associated with the coordinate $X^{\pm}$ compared to the case when the $C_{\pm}$ fluxes are set off. 
%
%

\section{Discussion and Conclusions}
In this paper we study the effects of fluxes in the quantum properties of the M2-brane. We discuss the Hamiltonian formulation of the M2-brane compactified on a torus with 2-form fluxes induced by the 3-form $C_\pm$. We study the effect of the fluxes on the quantum properties of the theory. We  establish an equivalence relation between a theory that contains 2-form fluxes induced by the $C_{\pm}$ on the target space 'M2-brane with fluxes' and the supermembrane satisfying a topological condition over the worldvolume  associated to an 'irreducible wrapping' where no reference to the 3-form background is present. 
When we consider the M2-brane on a $C_{\pm}$ background and the target space is noncompact  or even compactified on the 2-torus times Minkowski but no fluxes are present, the spectrum of the theory is continuous.  Classically it can be understood from the fact that it contains -as in the uncompactified case-, string-like spikes that can be attached to the spectrum without any cost of energy.
The case we analyze corresponds to have 2-form fluxes on the target space induced by the $C_{\pm}$. In this case the spectral behaviour of the theory changes drastically: its mass spectrum becomes discrete.  The flux backreacts on the worldvolume generating a induced flux on the worldvolume associated to the presence of fixed nontrivial $U(1)$ fiber bundle whose first chern class is $k$. It acts as a new  constraint on the Hamiltonian  and it is associated to the existence of a nontrivial central charge condition. The Hamiltonian becomes the Hamiltonian of the Supermembrane theory irreducible wrapped on a flat torus, shifted by a constant term proportional to $k$. An immediate consequence of this equivalence between both actions is the fact that a supermembrane formulated on a Minkowski background in the presence of a three form $C_{\pm}$ toroidally compactified with an induced 2-form flux condition has a purely discrete spectrum with eigenvalues of finite multiplicity at quantum level. It represents a new sector of M2-brane with this property.
The membrane momentum becomes shifted by the flux units in the directions of the $X^+$ or $X^-$ coordinates corresponding to fluxes $C_{+}$ or $C_{-}$ respectively.
An old question posed by the authors in \cite{connes}  was the relation between the matrix model on a noncommutative torus  and its M-theory origin in terms of a M2-brane in the presence  of a quantization condition over a constant $C_-$ \cite{dwpp3}. We show that this last theory corresponds exactly to the supermembrane with central charges.
The results shown in the paper suggest the interest to generalize  the precise relation between fluxes and quantization properties of M2-brane in order to describe new sectors of the microscopic degrees of freedom of M-theory. 

\section{Acknowledgements}  A.R. and M.P.G.M. are partially supported by Projects Fondecyt 1161192 (Chile). C.L.H Y P.L. are supported by the Project ANT1756, ANT1855 and ANT1856 of the Universidad de Antofagasta. J.M.P. is grateful to Universidad de Antofagasta for kind hospitality and financial support during part of the realization of this work supported by MINEDUC-UA project, code Projects ANT1756 and ANT1855. Also, J.M.P. want to thank to Professor P. Bargueno for his hospitality at Uniandes (Bogot\'a, Colombia) during a visiting research from August to September of 2018 where part of this work was also done.
%
%

%
%
%





\end{document}